\documentstyle[multicol,aps,prl,epsf]{revtex}

\draft

\begin{document}
\draft
\author{D.L. Zhou, and C. P. Sun $^{a,b}$}
\address{Institute of Theoretical Physics, Academia Sinica, \\
P.O.Box 2735, Beijing 100080, China}
\title{Second order quantum decoherence in the boson system}
\maketitle

\begin{abstract}
The second order quantum decoherence (SOQDC)is proposed as a novel
description for the loss of quantum coherence only reflected by second order
quantum correlations. By calculating the two-time correlation function, the
phenomenon of SOQDC is studied in details for a simple model, a two boson
system interacting with a reservoir composed of one or many bosons. The
second order quantum decoherence effects can be observed in the sketched
cavity QED experiment.
\end{abstract}

\pacs{PACS number(s): 03.65.Bz, 03.65.-w, 03.65.Sq}

\pagenumbering{roma}

\thispagestyle{empty}
\widetext

\pagenumbering{arabic}

\begin{multicols}{2}

Embodying the wave nature of particles in the quantum world, the coherence
of a quantum system is manifested in the observation of interference fringes
reflected both by the first order correlation functions and by higher order
ones introduced by Glauber many years ago \cite{Glauber}. In the presence of
external quantum system (i.e. a reservoir) interacting with the studied
system, the quantum decoherence of the system happens with the disappearance
of interference fringes. This decoherence mechanism provides essential
elements in the understanding of quantum measurement\cite{Wheeler} and the
transition from quantum to classical mechanics\cite{Zurek}. However, to our
best knowledge, in both aspects from experiments\cite
{Wineland1,Haroche,Rampe,Umansky,Wineland2} and theories\cite{Sun1,Sun2},
most studies focus on the first order quantum decoherence, in which the
quantum coherence is described by the superposition of two or several single
states and reflected by the first order correlation function . In this
Letter we will present a novel description of decoherence phenomenon
reflected by the second order quantum correlation functions.

Actually the higher order effect of quantum coherence is manifested in the
multi-particle picture \cite{Scully}. For instance, for a single particle,
the coherent superposition $|s\rangle =\frac 1{\sqrt{2}}(|k\rangle
+|-k\rangle )$ of two momentum states $|k\rangle $ and $|-k\rangle $ with
opposite wave vectors explicitly shows the first order quantum coherence
reflected by the interference fringes $I(x)=|\langle x|s\rangle |^2=\sin
^2(kx)$. However, there indeed exists such a situation e.g., a single two
particle state $|k,-k\rangle $, in which there is not the first order
quantum coherence, but we can see the second order effect through the second
order quantum correlation functions. This fact even enjoyed by
Hanbury-Brown-Twiss experiment \cite{Scully} is exactly necessary to building
quantum optics. Therefore,it is natural to extend the investigations of
decoherence to the case of second order and to probe how an external system
destroys the interference fringes due to high order coherence.

Studying the high order decoherence quantitatively without much technical
disturbance, we first restrict ourselves to a simple model: the two mode
bosonic system coupling to a reservoir composed of one or many bosons. In
the case of one reservoir boson being in the Fock state, the second order
decoherence factor measuring the loss of second order coherence is found to
be a fast oscillating function when the particle number becomes larger. In
the case of many reservoir bosons being in vacuum state, the second order
decoherence factor manifests the quantum revival-collapse phenomena of long
time. This kind of quantum jump even was predicted by one (CPS) of the
authors for the first order quantum decoherence with the universal
factorization structure\cite{Sun1}. In this letter it is found that, in the
continuous mode limit with some given spectral distributions, the second
order decoherence factor will decay following an exponential law. To test
this theoretical model, a Gedanken experiment is proposed: two bosonic atoms
pass through a one-mode cavity or many-mode (leaky) cavity. The possibility
of the realization of the experiment will be only briefly discussed. We take
$\hbar =1$ in this letter.

The model Hamiltonian $H=H_0+V$ is defined by
\begin{eqnarray}
H_0 &=&\omega _e\hat{b}_e^{\dagger }\hat{b}_e+\sum_j\omega _j\hat{a}%
_j^{\dagger }\hat{a}_j, \\
V &=&\sum_jd(\omega _j)(\hat{a}_j \hat{b}_e^{\dagger }\hat{b}_g+%
\hat{a}_j^{\dagger }\hat{b}_g^{\dagger }\hat{b}_e),
\end{eqnarray}
where $H_0$ is the free Hamiltonian of the system plus the reservoir, $V$
the interaction between the system and the reservoir and $\hat{b}_e^{\dagger
}(\hat{b}_e),\hat{b}_g^{\dagger }(\hat{b}_g)$ the creation (annihilation)
operators of two modes labeled by index $e$ and $g$ . Their frequencies are $%
\omega _e$ and $\omega _g=0$ respectively. The operators $\hat{a}_j^{\dagger
}(\hat{a}_j)$ are creation (annihilation) operators of the modes which
labeled by index $j$ and the frequency of each mode is denoted by $\omega _j$%
. The frequency-dependent constant $d(\omega _j)$ measures the coupling
constant between the system and the $j$ mode of the reservoir. This model
can be physically implemented as a cavity QED system, in which many
identical bosonic atoms of two levels ($|e\rangle $ and $|g\rangle $ are
placed in a cavity with some modes of frequencies $\omega _j$ . $H$ defines
the second quantization Hamiltonian for the atom-cavity system.

In our model the interaction is turned on only in the time interval $t\in
[0,T]$. This time period can be understood as the time of atom passage in
the cavity. The coupling with the reservoir-the multi-mode cavity field will
change the second order coherence of atoms. After time $T$, an ensemble
measurement in term of the second order correlation function is made on the
system so that the decoherence effect in the above time period can be
observed directly. The second order correlation functions
\begin{equation}
G[t,t^{\prime },\hat{\rho}_S(T)]=Tr_S(\hat{\rho}_S(T)\hat{B}^{\dagger }(t)%
\hat{B}^{\dagger }(t^{\prime })\hat{B}(t^{\prime })\hat{B}(t)),
\end{equation}
is defined as a functional of the the reduced density operator $\hat{\rho}%
_S(T)$ of the system for a given time $T$. Here, the atomic field operator
\begin{equation}
\hat{B}(t)=\exp (i\hat{H}_0t)[c_1\hat{b}_g+c_2\hat{b}_e]\exp (-i\hat{H}_0t)
\end{equation}
describes a specific quantum measurement with respect to the superposition state
$|+\rangle =c_1|e\rangle-c_2|g\rangle $ where $c_1$ and $c_2$ satisfy the
normalization relation $|c_1|^2+|c_2|^2=1$. Without loss of  generality, we
take $c_1=c_2=1/\sqrt{2}$ standing for a measurement in terms of $\hat{B}%
(t)=1/\sqrt{2}[\hat{b}_g+\hat{b}_e\exp (-i\omega _et)]$ as follows.

Before entering the main issue, let us consider what kind of atomic states
can demonstrate the second order quantum decoherence of the identical atom
system directly. To this end we temporarily analyze the case free of
interaction, i.e., we assume $d(\omega _j)=0$ for the moment. We denote the
Fock states for the two-level atom by $|n_g,n_e\rangle .$ Obviously, the two
particle states $|0_g,2_e\rangle ,|1_g,1_e\rangle $ and $|2_g,0_e\rangle $
span a 3-dimensional invariant subspace. With respect to the states $%
|0_g,2_e\rangle $ and $|2_g,0_e\rangle $, in each of which two particles
occupy the same position, the two second order correlation functions take
the same constant $\frac 12$ and thus do not show quantum interference.
However, a direct conclusion illustrates that only in the state $%
|1_g,1_e\rangle $, is the second order correlation function\cite{Tannoudji1}
$G(t,t^{\prime })=\cos ^2[\frac 12\omega _e(t-t^{\prime })]$  just an
interference term in time domain. The crucial point to be emphasized is that
for the general case with non zero coefficients the result of this
conclusion is invariant when any other measurement is taken on the system
under the condition that $c_1c_2\neq 0$. So, in the following discussions,
the second order decoherence is studied for the initial state  each
component of which possesses one particle. To purify the central idea, we
will distinguish the two cases with one mode reservoir and many mode
reservoir respectively.

In the one mode case, the whole system is initially prepared in the initial
state
\begin{equation}
|\psi (0)\rangle =|1_g,1_e,N\rangle ,
\end{equation}
It is easy to see that there is an invariant subspace $V^N$ with the basis $%
\{|0_g,2_e,(N-1)\rangle ,|1_g,1_e,N\rangle ,|2_g,0_e,(N+1)\rangle \}$. Since
these vectors are determined completely by the particle number in the
reservoir mode and thus a simple notation can be introduced without
confusion: $|N-1)\equiv |0_g,2_e,N-1\rangle $, $|N)\equiv |1_g,1_e,N\rangle $%
, $|N+1)\equiv |2_g,0_e,N+1\rangle $. Because the invariant subspace $V^N$
is also closed for the evolution operator $\hat{U}(T)=\exp (-i\hat{H}T)$,
the time evolution of the whole system can be determined as a sub-evolution $%
|\psi (T)\rangle =\hat{U}(T)|N)$ in $V^N$. Correspondingly, the reduced
density operator of the system can be calculated as
\begin{eqnarray}
\hat{\rho}_S(T) &=&|1_g,1_e\rangle \langle 1_g,1_e||(N|\hat{U}(T)|N)|^2
\nonumber \\
&\mbox{}&+|0_g,2_e\rangle \langle 0_g,2_e||(N-1|\hat{U}(T)|N)|^2  \nonumber
\\
&\mbox{}&+|2_g,0_e\rangle \langle 2_g,0_e||(N+1|\hat{U}(T)|N)|^2
\end{eqnarray}
By making use of the normalization condition
\begin{equation}
\sum_{M=N-1}^{N+1}|(M|\hat{U}(T)|N)|^2=1
\end{equation}
the second order correlation function can be calculated in a compact form
\begin{equation}
G[t,t^{\prime },\hat{\rho}_S(T)]=\frac 12[1+|(N|\hat{U}(T)|N)|^2\cos (\omega
_e(t-t^{\prime })].  \label{seccor}
\end{equation}
Obviously, the decoherence effect is completely characterized by the
decoherence factor $|(N|\hat{U}(T)|N) |^2$. When this factor is
equal to $1$, the interference fringe is clearest; As it becomes zero, the
interference fringe disappear completely.

Now, we adopt the resovolent method\cite{Tannoudji2} to calculate the
decoherence factor $(N|\hat{U}(T)|N) $. Its Fourier
transformation is written as
\begin{equation}
(N|\hat{U}(T)|N) =\frac i{2\pi }\int_{-\infty }^\infty
dze^{-izT}(N|\frac 1{z-\hat{H}+i0^{+}}|N) .
\end{equation}
A straightforward non-perturbative calculation gives the diagonal element
\begin{equation}
(N|\frac 1{z-\hat{H}}|N) =\frac 1{z-E_N-\frac{2Nd^2}{%
z-E_N+\Delta }-\frac{2(N+1)d^2}{z-E_N-\Delta }}.
\end{equation}
of the resovlent $G(z)=\frac 1{z-\hat{H}}$ for $E_N\equiv \omega _e+N\omega
_j$, $\Delta \equiv \omega _j-\omega _e$ and $d=d(\omega _j).$

In the resonant case, i.e. $\Delta =0$, there exist two first order poles of
$G(z)$ at $z=E_N\mp |d|\sqrt{4N+1}$ . Using the Cauchy's integral theorem,
the decoherence factor can be explicitly obtained as a oscillating function
\begin{equation}
|(N|\hat{U}(T)|N) |^2=\cos ^2[(4N+1)d^2T].
\end{equation}
of the passage time $T$ with the frequency $2(4N+1)d^2$. This perfect
oscillation is exactly the manifestation of the complete collapse- revival
of the second order coherence, which is very similar to that in the first
order quantum decoherence.

For the off-resonant case $\Delta \neq 0$, the analytical result for $%
|(N|\hat{U}(T)|N) |^2$ can also be obtained by the same method,
but it will be verbose to demonstrate the physical meaning explicitly.
Therefore, we numerically calculate the decoherence factor in FIG. $1$ for $%
N=0$, $N=1$, $N=2$, $N=7$ respectively.

\begin{figure}[btp]
\begin{tabbing}
\= \epsfxsize=4.15cm\epsffile{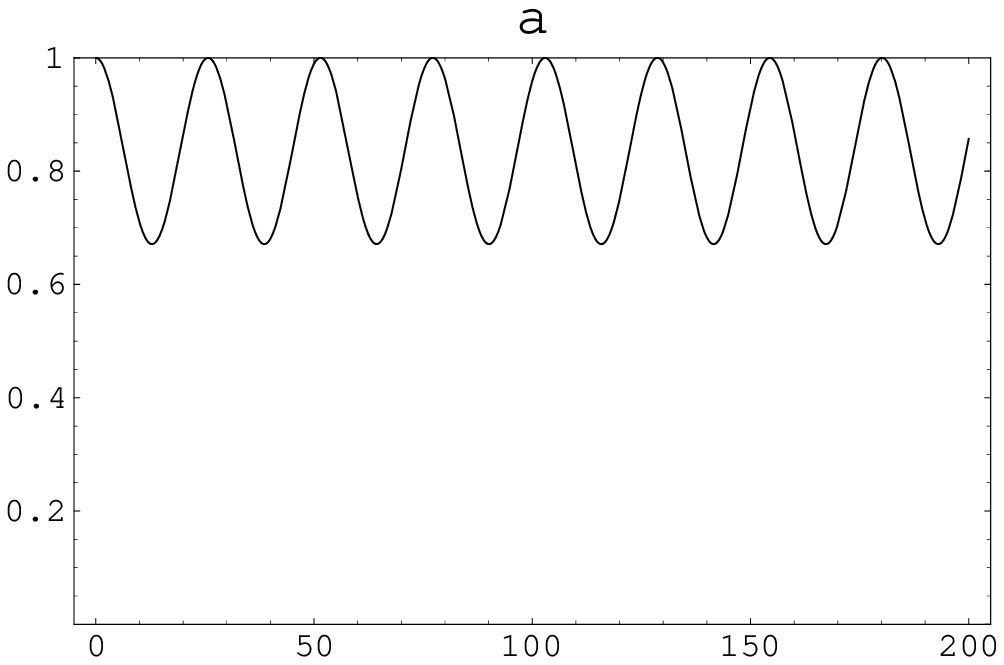}
\= \epsfxsize=4.15cm\epsffile{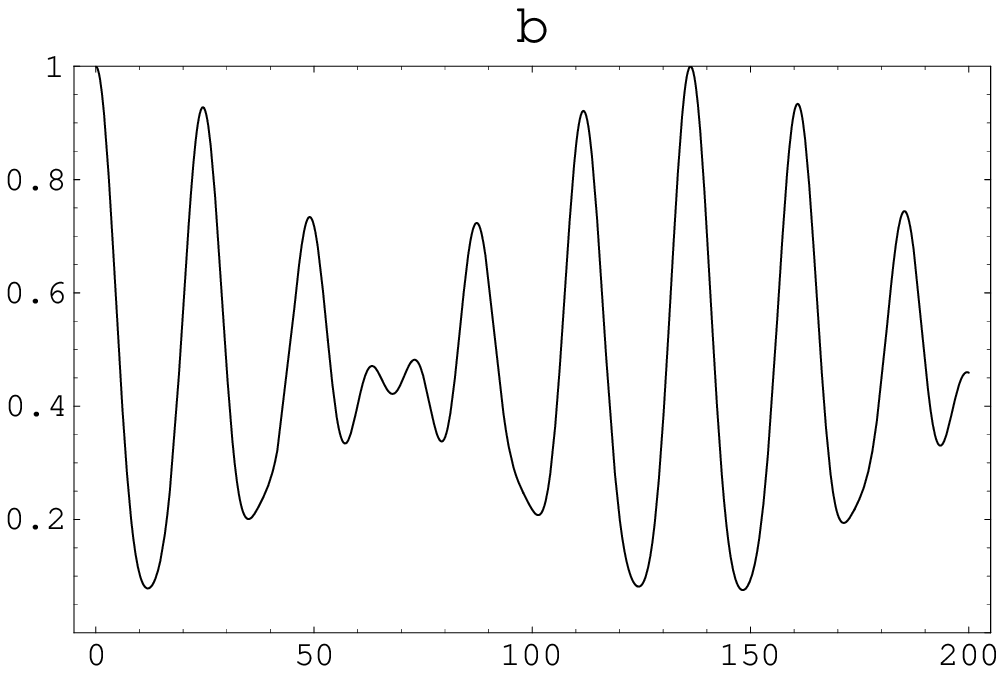}\\
\> \epsfxsize=4.15cm\epsffile{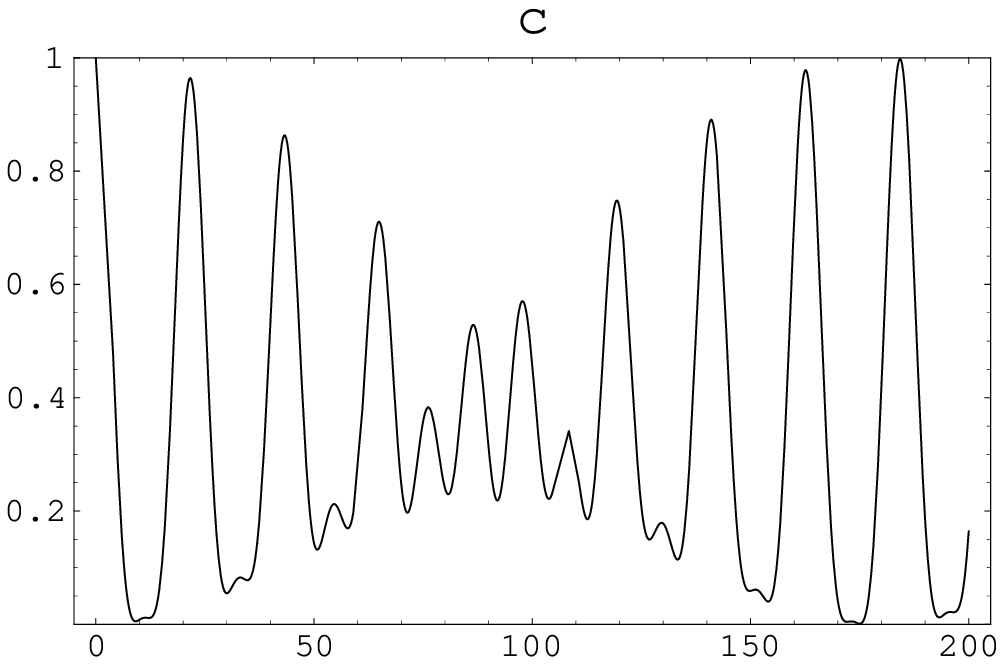}
\> \epsfxsize=4.15cm\epsffile{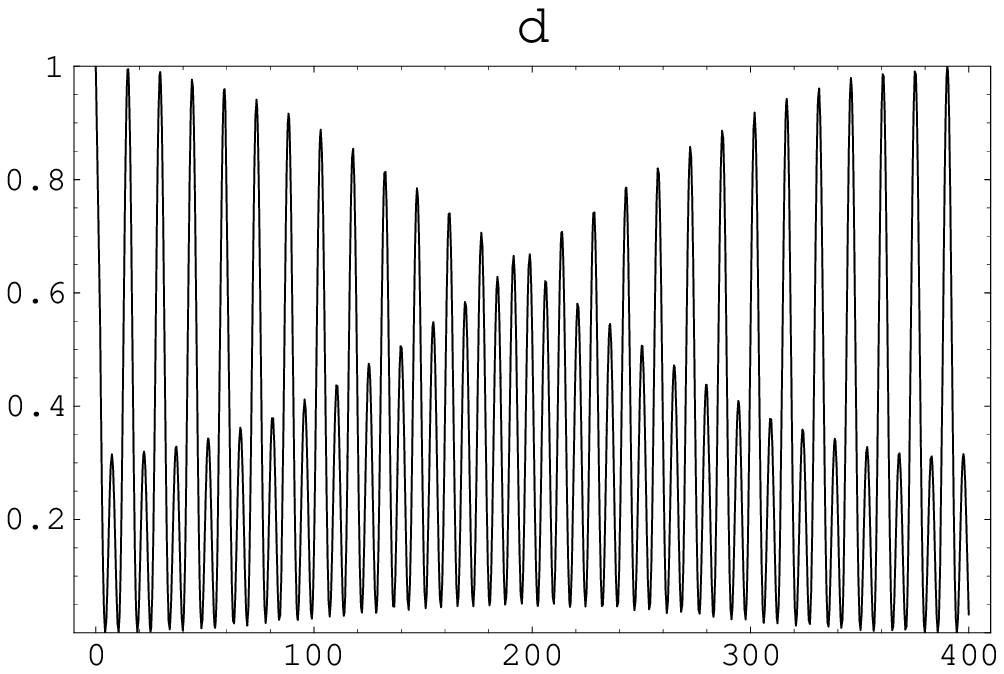}
\end{tabbing}
\caption{The horizontal axe denotes time period $T$, the vertical axe
denotes the decoherence factor $|(N|\hat{U}(T)|N)|^2$,
parameters $\Delta=0.20$, $d=0.07$, (a)$N=0$, (b)$N=1$, (c)$N=2$, (d)$N=7$.}
\end{figure}

From FIG. $1$, we observe that when one mode reservoir is just in the vacuum
state, the decoherence factor makes a sinusoidal oscillation. As the photon
number $N$ in the cavity becomes larger and larger, the decoherence factor
oscillates faster and faster regularly. From the point of view of the second
order quantum decoherence , this numerical result without decaying of
coherence re-proves that the Fock state of photon is basically quantum even
for very large photon number $N$ forming a macroscopic quantum state. Unlike
the quasi-classical state or a factorized macroscopic state, e.g., the
coherent state, it can not decohere its coupling quantum system into a
classical one.

To examine whether the macroscopic feature of the reservoir or the cavity
causes the second order decoherence or not , we consider the many mode
cavity -atom system in an initial state $|\psi (0)\rangle =|1_g,1_e\rangle
\otimes |\{0_j\}\rangle $ where $|\{0_j\}\rangle $ is the vacuum state of
the cavity .We denote the general Fock states of the many mode field by $%
|\{n_j\}\rangle \equiv $ $|n_1,n_2,...\rangle $.

Similar to the case with single mode cavity, there is still an invariant
subspace spanned by $\{|1_g,1_e,\{0_j\}\rangle ,|2_g,0_e,1_j,\{0_m\}_{m\neq
j}\rangle \}$ conserving the total atomic number. These basis vectors are
also determined completely by the particle number in one reservoir mode and
its index, and  similar notation will be adopted for them: $|0)\equiv
|1_g,1_e,\{0_j\}\rangle $, $|1_j)\equiv |2_g,0_e,1_j,\{0_m\}_{m\neq
j}\rangle $. As in the case of one mode reservoir, we can verify that the
second order correlation function is still in the same form (\ref{seccor})
except that $|(N|\hat{U}(T)|N)|^2$ is replaced by $|(0|\hat{U}(T)|0)|^2$.
Obviously, the decoherence effect is similarly determined by the second
order decoherence factor $|(0|\hat{U}(T)|0)|^2$ completely.

In this case, the resovlent has the expectation value
\begin{equation}
(0|G(z)|0)=\frac 1{z-\omega _e-\sum_j\frac{2d^2(\omega _j)}{z-\omega _j}}.
\end{equation}
In the limit of continuous mode, the Wigner-Weiskoff approximation\cite
{Louisell},
\begin{equation}
\sum_j\frac{2d^2(\omega _j)}{z-\omega _j+i0^{+}}=\Delta _e-i\frac{\Gamma _e}2%
,
\end{equation}
deduces the exponential decaying decoherent factor
\begin{equation}
|(0|\hat{U}(T)|0) |^2=\exp (-\Gamma _eT).
\end{equation}
for the decaying rate $\Gamma _e=4\pi \rho (\omega _e)d^2(\omega _e)$ , the
renormalization of frequency
\begin{equation}
\Delta _e=2P\int d\omega _j\rho (\omega _j)\frac{d^2(\omega _j)}{\omega
_e-\omega _j},
\end{equation}
and $\rho (\omega _j)$ is the mode density of the reservoir. Obviously, it
represents an irreversible decay of the second order coherence. Compared
with the usual result of the spontaneous radiation decay rate for one
excited particle, this decay rate $\Gamma _e$ is two times of that which
manifests the bosonic stimulation effect, that is, the existence of one
boson in the ground state increases the transition amplitude of the other
boson from the excited state to the ground one.

In fact, the continuous mode limit is an extreme situation with infinite
modes of reservoir. Like the studies for the first order quantum coherence
in our previous work\cite{Sun1,Sun2}, it is very interesting to consider how
the quantum system graduates into the second order decoherence as the mode
number of the coupled external system becomes larger step by step. This
problem possibly reveals the physical process extrapolating quantum to
classical in the case that there does not exist the first order coherence,
but there exists the second order coherence. In our model we study the case
that the number of the modes of the reservoir is finite. The numerical
results of the decoherence factors $|(0|\hat{U}(T)|0) |^2$ are
given in FIG. $2$ for different numbers of the modes of the reservoir. With
the number of the modes increasing, the quantum revival-collapse phenomena
in the decoherence factor is clearly illustrated in FIG. $2$. It should be
emphasized that the decoherence rate becomes shorter and the quantum revival
time becomes longer when the mode density increases. This implies the
irreversibility of the decoherence process when the modes become continuous. %
\begin{figure}[btp]
\begin{tabbing}
\= \epsfxsize=4.15cm\epsffile{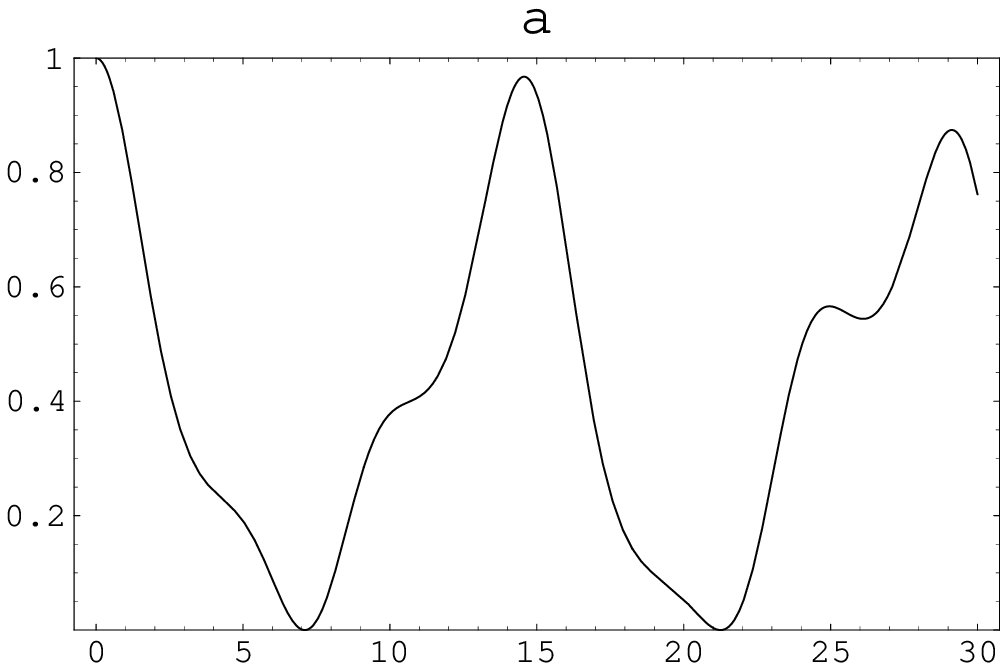}
\= \epsfxsize=4.15cm\epsffile{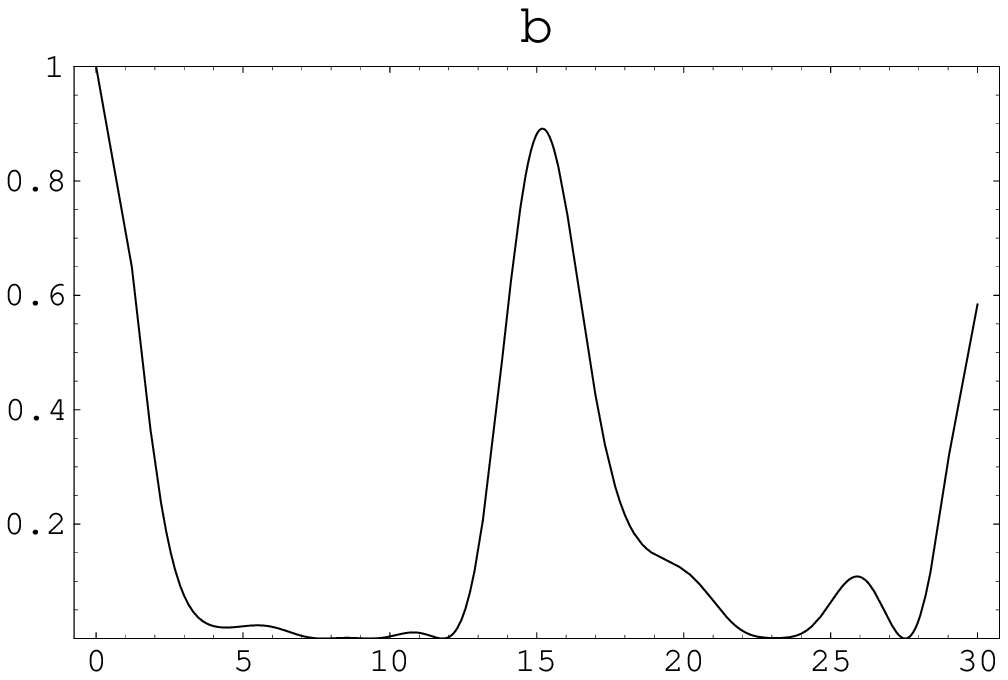}\\
\> \epsfxsize=4.15cm\epsffile{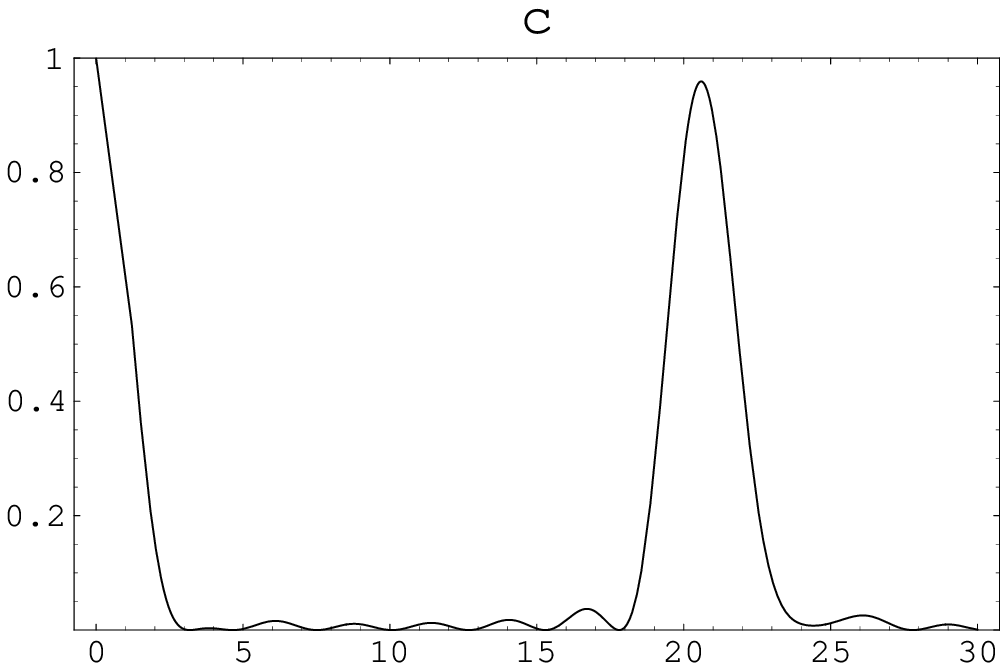}
\> \epsfxsize=4.15cm\epsffile{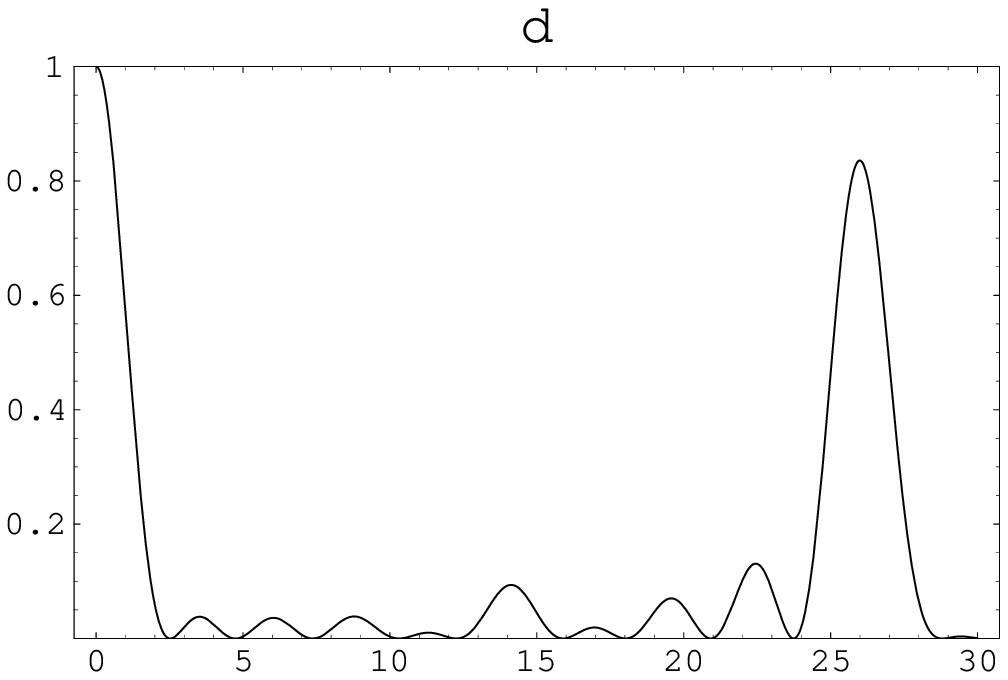}
\end{tabbing}
\caption{The horizontal axe denotes time period $T$, the vertical axe
denotes the decoherent factor $|(0|\hat{U}(T)|0) |^2$,
parameter $d(\omega _j)=0.17$, the number of the modes of the reservoir is
different: (a)$N_{mod}=3$, (b)$N_{mod}=5$, (c)$N_{mod}=7$, (d)$N_{mod}=9$.}
\end{figure}

Before concluding this letter, we sketchily discuss the possibility to
observe this second order decoherence process in a cavity QED experiment.
The experiment is arranged as follows in three steps. The first step is to
prepare the bosonic atoms in the initial state $|1_g,1_e\rangle $ while the
cavity is in an appropriated state. Then, we control the two atoms passing
through the cavity within a time period $T$, which is determined by the
velocity of the atoms. In the third step, to show the second order
coherence, a coincidence measurement for the two atoms in two different
locations is made by two detectors at t and t'. The sketched experiment
setup is illustrated in FIG. $3$. One of the possible difficulties in
experiments is the preparation of the initial quantum state of the two atoms
and the cavity field. With better controlling of the atomic motion and the
state of the cavity field in experiments\cite{Experiments}, we believe, this
is not a big problem in the near future.

\begin{figure}[btp]
\centerline{\epsfxsize=6.6cm\epsffile{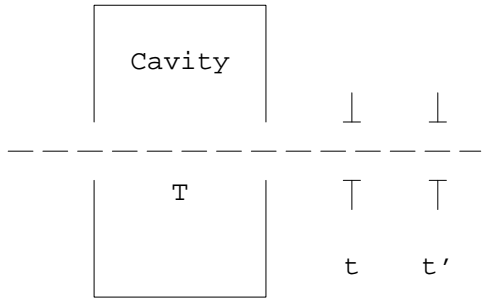}}
\caption{Two bosonic atoms pass through a cavity in a time period $T$ and
are detected coincidentally by two detectors in different time $t$ and $%
t^{\prime }$.}
\end{figure}

\vskip 0.2cm In sum, we have proposed the dynamic description of the loss of
the second order coherence by studying the simplest case : the considered
system is composed of  two bosons in two modes and the reservoir composed of
one or many bosons, and we have suggested a possible cavity QED experiment
to detect this process of the second order quantum decoherence.

\vskip 0.2cm {\bf Acknowledgement} One of the authors (D.L. Zhou) is
grateful to B.F. Zhu and Z. Xu for helpful comments. And we are also grateful
to X.F. Liu for reading the paper carefully. This work is supported
by NSF of China. 

\end{multicols}

\end{document}